\documentclass[letterpaper]{article} 
\usepackage[preprint]{aaai2027}
\usepackage[hyphens]{url} 
\usepackage{graphicx} 
\usepackage{natbib} 
\usepackage{caption} 
\usepackage{amsmath,amssymb}
\usepackage{booktabs}
\usepackage{array}

\graphicspath{{figures/paper/}}

\newcommand{\method}{\textsc{MoGI}}
\newcommand{\R}{\mathbb{R}}
\newcommand{\cG}{\mathcal{G}}
\newcommand{\cV}{\mathcal{V}}
\newcommand{\cE}{\mathcal{E}}
\newcommand{\cM}{\mathcal{M}}
\newcommand{\cT}{\mathcal{T}}
\newcommand{\CE}{\operatorname{CE}}

\newcommand{\givetabhead}[1]{\textbf{#1}}
\newcommand{\givevalue}[1]{#1}
\newcommand{\givebestvalue}[1]{\underline{#1}}

\newcommand{\giveWideTableSetup}{\small\setlength{\tabcolsep}{1.6pt}\renewcommand{\arraystretch}{0.98}}

\newcommand{\giveCompactTableSetup}{\small\setlength{\tabcolsep}{2.2pt}\renewcommand{\arraystretch}{1.08}}

\title{Measuring the Influence of Modality-Specific Node Representations on Graphs}
\author{
Xu Wang\textsuperscript{\rm 1},
Xunkai Li\textsuperscript{\rm 2},
Yinlin Zhu\textsuperscript{\rm 3},
Rong-Hua Li\textsuperscript{\rm 2}
}
\affiliations{
\textsuperscript{\rm 1}School of Airspace Science and Engineering, Shandong University, WeiHai, China\\
\textsuperscript{\rm 2}Department of Computer Science, Beijing Institute of Technology, Beijing, China\\
\textsuperscript{\rm 3}School of Computer Science and Engineering, Sun Yat-sen University, GuangZhou, China\\
\textbf{Correspondence to:} Rong-Hua Li \texttt{<lironghuabit@126.com>}
}

\begin{document}
\maketitle

\begin{abstract}
Multimodal attributed graphs combine graph structure with node text and image representations in recommendation and multimedia analysis. Graph predictors propagate these representations across connected nodes, so one input can change predictions at its own and nearby nodes. Measuring this distributed response reveals which modality-specific inputs a trained model relies on and how predictor design changes that reliance. We study influence measurement for modality-specific node representations. Given an encoded graph, a trained predictor, a training split, and fixed candidate representations, the task outputs one influence score per candidate and separate text and image rankings. We score a candidate by replacing it with its modality's training-node mean and comparing the original and changed predictions. Existing feature attribution and data valuation generally target one output, training examples, whole nodes, or graph elements rather than prediction changes across graph nodes. Our empirical study shows that nearby-node responses contain information missed by source-only scores and simple proxies, while response magnitude alone does not identify which class changes support the task. Guided by these observations, we propose \method{} (Modality-specific Graph Influence). It propagates training labels to define task direction, reads the same replacement response for the favored class and full class distribution, and averages the resulting ranks. \method{} requires no predictor retraining or learned scoring parameters. Under a shared five-dataset, eight-predictor benchmark, it achieves the largest aggregate held-out loss increase against all five external methods at every tested ranking cutoff.
\end{abstract}

\section{Introduction}

Multimodal attributed graphs (MAGs) combine relational structure with node inputs such as text and images, supporting applications including personalized recommendation and multimedia analysis~\cite{MultimodalGraphs,MultimodalGraphSurvey,MMGCN,MGAT}. Consider a product graph whose nodes carry description and photograph vectors. A graph predictor propagates these inputs between connected products, so changing one representation can alter predictions at both the node carrying it---the source node---and nearby nodes. This distributed response makes it difficult to determine which text or image representations most strongly shape the model's predictions. Measuring influence at this granularity reveals where the predictor relies on each modality and how its response spreads across connected nodes. Comparing such scores across predictors shows how encoder, fusion, and propagation choices change the representations to which a model is most responsive, providing a concrete diagnostic for architecture analysis. Scoring an entire product as one unit would hide the different roles of its text and image representations, so we analyze the two modalities separately.

The same representation-level view supports model inspection when a prediction warrants closer review: it identifies the specific text or image inputs associated with large prediction changes instead of treating a multimodal node as an indivisible unit. We therefore study \emph{influence measurement for modality-specific node representations}. The task begins after the modality inputs and graph predictor have been trained. Its inputs are the encoded graph, the fixed predictor, the training split, a fixed set of candidate text and image vectors, and one training-mean reference---the average training-node vector---for each modality. Its primary output is one influence score for every candidate. To compute a score, we replace only that candidate by its same-modality reference, recompute predictions at the source and nearby nodes, and then restore the original input. Repeating this operation produces complete and separate text and image rankings. Together, the scores form a diagnostic map of which modality-specific inputs affect the predictor and whether their responses remain local or spread to neighboring nodes.

This setting is easily confused with several neighboring tasks, but its decision and output are different. Active acquisition and missing-modality completion decide which absent inputs to obtain or reconstruct~\cite{CAMA}; model compression changes the representations or predictor retained for deployment; graph explainers typically explain one chosen prediction; and data valuation assesses training examples or graph elements. In contrast, our graph, candidate set, encoded representations, and predictor are fixed, no modality is acquired or deleted, and the output is a complete diagnostic ranking. The 5\%, 10\%, and 20\% prefixes used later are evaluation cutoffs rather than resource budgets. Moreover, replacements are not accumulated: every candidate is scored from the same full input, so the task is neither sequential selection nor joint-subset optimization.

Feature attribution and graph explanation provide established tools for relating input features, edges, or subgraphs to one chosen prediction~\cite{IntegratedGradients,GNNExplainer,GraphMask,SubgraphX,GraphFramEx}. Graph data valuation provides a complementary view by assigning value to training examples, whole nodes, graph elements, or inference neighbors~\cite{InfluenceFunctions,DataShapley,GraphElementInfluence,PCWinter,NORA,SGUL,NodeGraphValuation}. This representation-level analysis asks a different question: which already encoded text or image representations produce prediction changes that are most consequential for the task? A response-magnitude score remains class-agnostic: a large probability change can move mass among classes without indicating whether the affected representation supports the task. Held-out labels could reveal that direction, but using them to construct the score would compromise evaluation.

Two empirical questions determine how to measure the graph response. Does source-only scoring miss useful changes at nearby nodes, and does direct probability change reveal more than statistics such as degree, confidence, and uncertainty? Expanding the scope to the one-hop and two-hop neighborhood---nodes one or two graph edges away---improves the ranking at every cutoff, while direct change beats the strongest tested proxy in 13 of 15 dataset--seed groups. These findings still leave the class-agnostic response without task direction. Before opening the test results, validation on predeclared development seeds therefore asks whether a class distribution formed only from propagated training labels can supply that direction without using validation or test labels during scoring. Combining two training-label-directed readings improves over the best-performing validation comparator on all five datasets and most predictors at every cutoff.

Motivated by these findings, we propose \method{}. It measures one task-directed graph response for each candidate: how replacing its node-specific information by a same-modality training mean changes support for task-relevant classes at the source and nearby nodes. The replacement defines the comparison, the graph targets define where the response is observed, and a fixed propagation of training labels supplies its task direction. The class favored by this direction and the full class distribution provide two readings of the same prediction change. Averaging their candidate ranks produces one deterministic ranking per modality. The predictor remains frozen, and the same procedure applies to any trained graph predictor that returns class probabilities.

\noindent\textbf{Our Contributions:}
(1) \textbf{Valuable Insights.} We formulate influence measurement for modality-specific node representations in frozen graph predictors. The resulting scores separate text from image and reveal whether each representation's response remains at its source or spreads to nearby nodes.
(2) \textbf{Novel Method.} We introduce \method{}, which measures one task-directed graph response through a common reference comparison, training-label direction, and two readings of the same prediction change. It requires no predictor retraining, adds no learned parameters, and reuses one replacement forward pass.
(3) \textbf{State-of-the-art Performance.} Under a shared five-dataset, eight-predictor protocol, \method{} attains the largest aggregate held-out loss increase against all five external methods at the 5\%, 10\%, and 20\% cutoffs. The ranking transfers across all eight predictors, and its margin over the next-best aggregate method has a positive 95\% interval at every cutoff.

\section{Related Works}

\textbf{Multimodal graph learning.}
MAG predictors fuse modality inputs through graph convolution, attention, recommendation-specific propagation, graph filtering, cross-modal interaction, or local--global aggregation~\cite{MMGCN,MGAT,LGMRec,MIGGT,Graph4MM,DMGC,DGF}. Recent work further studies modality-aware geometry and roles, missing-modality completion, aligned propagation, co-evolution, clustering, pretraining, self-teaching, generative learning, unified encoders, and multimodal graph assistants~\cite{GraphMLLM,UniGraph2,LION,OpenMAG,NTSFormer,TMTE,RoleMAG,CAMPA,CoMAG,GREMC,SMGFM,MMGL,GraphGPTO,MLaGA,GraphMNL}. General cross-modal encoders provide another source of text and image representations~\cite{CLIP,ImageBind}. These methods learn graph predictors from multimodal inputs, whereas \method{} analyzes how individual encoded text or image representations shape the predictions of a trained model.

\textbf{Explanation and valuation.}
Feature and graph explainers attribute a chosen prediction through coalitions, paths, masks, perturbations, analytical decompositions, or subgraphs~\cite{IntegratedGradients,GNNExplainer,Captum,PGExplainer,GraphMask,ShapleyFlow,SubgraphX,GraphSVX,PGMExplainer,CFGNNExplainer,ProbingGNNExplainers,GraphFramEx,GNNShap,GOAt}. Data- and graph-valuation methods target training examples, datasets, nodes, edges, coalitions, or test-time neighbors~\cite{InfluenceFunctions,DataShapley,DistributionalDataValuation,BetaShapley,DataShapleySelection,EfficientDataShapley,TracIn,DVRL,PCWinter,GraphElementInfluence,NORA,SGUL,NodeGraphValuation}. We adapt representative explainers to the node--modality unit---one text or image vector at one node---and aggregate their model response over the source and nearby graph nodes. Active acquisition obtains missing modalities~\cite{CAMA}, while \method{} ranks already encoded representations by their task-directed graph response.

\begin{figure*}[!t]
\centering
\includegraphics[width=0.94\textwidth]{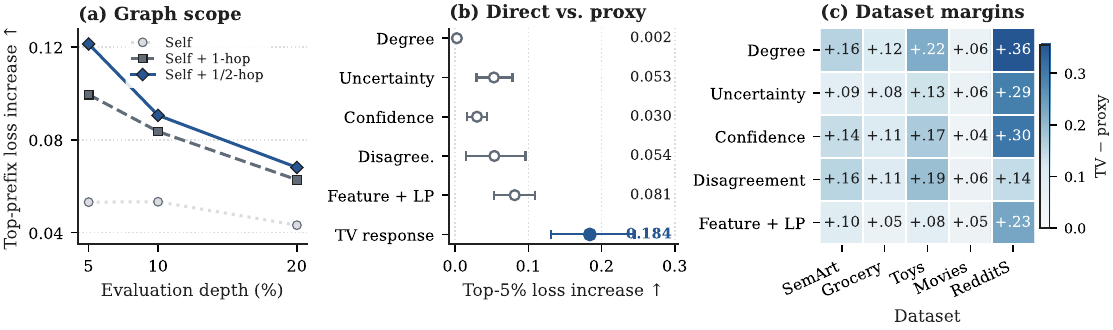}
\caption{\textbf{Graph scope and direct response.} (a) Adding one- and two-hop targets improves loss increase at every depth. (b) Direct total-variation (TV) response exceeds structural and predictive-state proxies on average. (c) Its advantage remains positive on all five datasets; intervals resample dataset--seed groups, and Feature + LP combines fused-feature and label-propagation probabilities. Together, the results motivate direct neighborhood-response measurement.}
\label{fig:study}
\end{figure*}

\section{Problem Formulation}
\label{sec:problem}

We formalize influence measurement for modality-specific node representations independently of any particular scoring method.

\textbf{Inputs, fixed state, and decision unit.}
Let $\cG=(\cV,\cE)$ be an encoded graph with adjacency $A$, class set $\{1,\ldots,C\}$, and modalities $\cM$. Node $i$ has a representation $h_i^m\in\R^{d_m}$ for modality $m$, where $d_m$ is its representation dimension; for example, $h_i^{\mathrm{text}}$ is its text vector. Let $H$ collect all node--modality representations. A trained predictor $f_\theta$ maps $A$ and $H$ to the original class probabilities
\begin{equation}
P=f_\theta(A,H)\in[0,1]^{|\cV|\times C}.
\label{eq:predictor-field}
\end{equation}
A candidate $e=(i,m)$ is one encoded representation: modality $m$ at node $i$. For example, $(i,\mathrm{image})$ denotes the photograph vector of product $i$, not the whole node, raw image, or one embedding coordinate. The task receives a training split, fixed candidate sets $\mathcal C_m$, one training-node mean reference per modality, and a target set $\cV_{\mathrm{tar}}\subseteq\cV$ whose predictions may enter a score. The graph, representations, candidates, and predictor remain fixed. Scoring may use the training split but not validation or test labels. It outputs one influence score $s(e)$ per candidate and a complete ranking $\pi_m$ within each modality.

\textbf{One-at-a-time reference response.}
The reference $r^m$ is the mean representation of modality $m$ over training nodes. For candidate $e=(i,m)$, we replace only $h_i^m$ by $r^m$, obtaining $H^{-e}$ and $P^{-e}=f_\theta(A,H^{-e})$. Every candidate starts from the same full input $H$ and original predictions $P$; the original input is restored before the next candidate is examined. Targets are grouped as the source itself, one-hop nodes, and two-hop nodes. When a distance group is large, a deterministic cap bounds its contribution to per-candidate scoring cost without changing the candidate set. We call the resulting frozen-predictor response the candidate's \emph{influence}; its score estimates this response without held-out labels. Text and image candidates are ranked separately because their representation spaces and score distributions need not be comparable.

\textbf{Held-out ranking evaluation.}
Let $\mathcal E_i$ be the held-out labeled targets reached by candidate $e=(i,m)$, let $y_j$ be target $j$'s label, and let $\CE$ denote cross-entropy. We set $u(e)=0$ when $\mathcal E_i=\varnothing$; otherwise,
\begin{equation}
u(e)=\frac{1}{|\mathcal E_i|}\sum_{j\in\mathcal E_i}
\left[\CE(P_j^{-e},y_j)-\CE(P_j,y_j)\right].
\label{eq:utility}
\end{equation}
Positive $u(e)$ means that replacement worsens held-out loss on average. Held-out labels appear only after every score and ranking has been finalized. At fractional depth $b$, we take the highest-scored $K_m=\min\{|\mathcal C_m|,\max(5,\lceil b|\mathcal C_m|\rceil)\}$ candidates and average $u(e)$; tables and figures call this quantity \emph{top-prefix loss increase}. The tested depths inspect the front of a complete ranking rather than limit how many candidates are scored. Because every replacement is independent, the task remains a static diagnostic rather than active acquisition, compression, or sequential subset selection.

\section{Empirical Study}
\label{sec:study}

Graph propagation raises two questions for representation ranking. \textbf{Q1:} Does a source-only score miss useful prediction changes at nearby nodes? \textbf{Q2:} Does direct probability change identify influential representations more effectively than simple scores based on graph structure or prediction probabilities?

\textbf{Study design.}
We use five multimodal graphs, three seeds, and separate text and image rankings. Each scoring rule orders candidates without evaluation labels. We then measure the mean increase in held-out cross-entropy when top-ranked representations are replaced one at a time; a larger increase means that representations supporting correct predictions appear earlier in the ranking. Q1 compares three nested prediction scopes on the same 220 candidates used by the main benchmark. Q2 uses a separate diagnostic sample spanning high-degree, uncertain, disagreement, random, and underrepresented-class cases, and compares direct probability change with five structural or predictive-state proxies at the 5\% depth. Its evaluation first averages loss within each graph-distance group and then weights groups by their size; sampling details appear in the Supplementary Material.

\textbf{Including nearby predictions yields more informative rankings (Answer for Q1).}
The complete self/one-hop/two-hop scope achieves the largest mean top-prefix loss increase at every depth (Fig.~\ref{fig:study}(a)). At 5\%, it improves over the source-only score by 0.0682 and over self plus one-hop by 0.0218, with clustered intervals $[0.0426,0.0932]$ and $[0.0072,0.0394]$. Thus, influence scoring should include nearby predictions rather than observe only the source node.

\textbf{Direct probability change outperforms the tested proxies (Answer for Q2).}
After averaging text and image results within each dataset--seed cluster, direct prediction change beats the strongest proxy in 13 of 15 clusters. Its mean clustered margin has a bootstrap 95\% interval of $[0.0238,0.0762]$ (Fig.~\ref{fig:study}), and no single proxy is strongest on every dataset. In this diagnostic, measuring the predictor's probability response is more informative than inferring influence from one structural or predictive-state statistic.

\textbf{Implication for Method Design.}
Together, the two findings define the quantity that \method{} must measure: the task-directed graph response of one representation under a common reference comparison. Direct response identifies where predictions move, while a class distribution formed only from propagated training labels identifies which class changes matter. On independent validation seeds, averaging the within-context ranks of the favored-class and full-distribution readings improves over the best-performing validation comparator on all five datasets and seven of eight predictors at every depth. This equal-rank rule is fixed before test evaluation; complete constituent results appear in the Supplementary Material.

\section{Methodology}
\label{sec:method}

\subsection{Overview}

\begin{figure*}[t]
\centering
\includegraphics[width=0.96\textwidth]{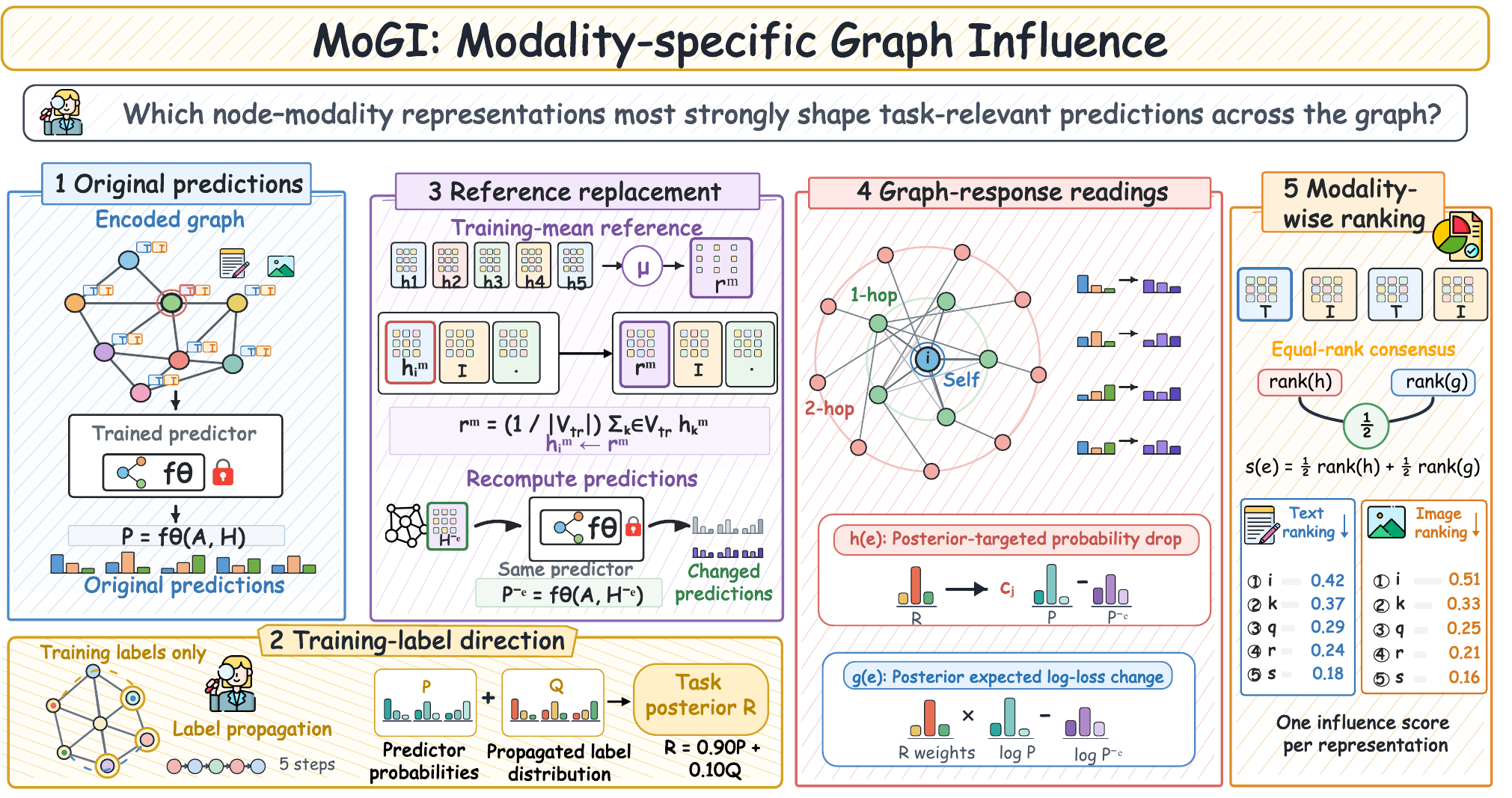}
\caption{\textbf{The \method{} pipeline.} A training-label posterior supplies task direction to the prediction change caused by replacing one node--modality representation with its training-mean reference. Favored-class and full-distribution readings use the same original and changed predictions over mutually exclusive self, one-hop, and two-hop targets; their equal-rank consensus produces separate text and image rankings.}
\label{fig:framework}
\end{figure*}

As illustrated in Fig.~\ref{fig:framework}, \method{} measures one quantity: how replacing the node-specific information in candidate $e=(i,m)$ changes support for task-relevant classes at the source and nearby nodes. The original and reference-replaced predictions define this response, while a training-label posterior---a class distribution obtained by propagating observed training labels across the graph---supplies its direction without held-out labels. One reading tracks the favored-class probability drop, and the other tracks expected log-loss change over the full class distribution. Equal-rank consensus retains candidates that rank highly under both readings without mixing their raw scales, producing one frozen-predictor ranking within each modality.

\subsection{Reference Comparison}

\textbf{Motivation.}
To isolate the prediction change associated with one representation, we replace one candidate at a time. For modality $m$, let $\cV_{\mathrm{tr}}$ denote the training nodes. We compute the training-mean reference
\begin{equation}
r^m=\frac{1}{|\cV_{\mathrm{tr}}|}\sum_{i\in\cV_{\mathrm{tr}}}h_i^m.
\label{eq:reference}
\end{equation}
For $e=(i,m)$, the reference-replaced input $H^{-e}$ replaces $h_i^m$ by $r^m$ and leaves every other representation unchanged:
\begin{equation}
H^{-e}=H\setminus\{h_i^m\}\cup\{r^m\},\qquad
P^{-e}=f_\theta(A,H^{-e}).
\label{eq:replacement}
\end{equation}
This reference provides a common same-modality comparison point. The resulting pair $(P,P^{-e})$ isolates the response associated with candidate $e$; every later step interprets this same prediction change.

\subsection{Training-Label Direction}

\textbf{Motivation.}
The replacement comparison shows how probabilities move, but magnitude alone does not identify which class changes matter for the task. The trained predictor describes its current belief, while graph-connected training labels provide task direction without exposing held-out outcomes. Let $Y\in\{0,1\}^{|\cV|\times C}$ contain one-hot labels on $\cV_{\mathrm{tr}}$ and zeros elsewhere. Starting from $Q^{(0)}=Y$, we apply $S$ fixed propagation steps with row-normalized adjacency $\widetilde A$:
\begin{equation}
\begin{aligned}
\overline Q^{(t+1)}&=\alpha Y+(1-\alpha)\widetilde A Q^{(t)},\\
Q_i^{(t+1)}&=
\begin{cases}
Y_i,&i\in\cV_{\mathrm{tr}},\\
\overline Q_i^{(t+1)},&i\notin\cV_{\mathrm{tr}},
\end{cases}
\end{aligned}
\label{eq:label-propagation}
\end{equation}
where $\overline Q^{(t+1)}$ is the provisional update before training rows are reset, and $\alpha$ controls the retained training-label mass. We normalize the final rows to obtain the propagated class distribution $Q$ and use a uniform row only when no propagated mass reaches a node. Resetting each training row to its observed one-hot label at every step keeps that row \emph{clamped}. For each target $j$, \method{} combines $Q_j$ with the frozen predictor probabilities $P_j$:
\begin{equation}
R_j=(1-\beta)P_j+\beta Q_j.
\label{eq:posterior}
\end{equation}
Here $R_j$ is the task-directed class distribution used for scoring, and $\beta$ is the fixed weight on propagated training-label information. The shared settings for $S$, $\alpha$, and $\beta$ are documented and justified in the Supplementary Material.

\subsection{Graph-Response Readings}

With the comparison $(P,P^{-e})$ and task direction $R$ fixed, \method{} reads one response over the source and nearby graph nodes. We divide the targets into three nonoverlapping groups: $\cT_{i,0}=\{i\}\cap\cV_{\mathrm{tar}}$ for the source, $\cT_{i,1}$ for targets at shortest-path distance one, and $\cT_{i,2}$ for targets at shortest-path distance two. Each node enters only its nearest group, and a deterministic rule retains at most $L$ targets from each nonempty group. With $\cT_i=\cT_{i,0}\mathbin{\dot\cup}\cT_{i,1}\mathbin{\dot\cup}\cT_{i,2}$, let $c_j=\arg\max_c R_j(c)$ be the class favored by $R_j$. The favored-class reading is the posterior-targeted probability drop
\begin{equation}
h(e)=\frac{1}{|\cT_i|}\sum_{j\in\cT_i}
\left[P_j(c_j)-P_j^{-e}(c_j)\right]
\label{eq:hard-score}
\end{equation}
and the posterior expected-log-loss change
\begin{equation}
g(e)=\frac{1}{|\cT_i|}\sum_{j\in\cT_i}\sum_{c=1}^{C}
R_j(c)\left[\log P_j(c)-\log P_j^{-e}(c)\right].
\label{eq:soft-score}
\end{equation}
The favored-class reading emphasizes the class selected by the posterior, whereas the full-distribution reading retains all posterior class probabilities. They are two views of the same task-directed graph response: both use the same $P$, $P^{-e}$, $R$, and target set, and require only one replacement forward.

\subsection{Consensus Ranking}

The two readings have different numerical scales, so \method{} combines their within-context orderings rather than their raw values. Within one dataset--split--predictor--modality context, let $N=|\mathcal C_m|$. Let $\operatorname{rank}_h(e)$ and $\operatorname{rank}_g(e)$ be average-tie ranks: larger scores receive larger ranks, and tied candidates receive the mean of the rank positions they occupy. Figure~\ref{fig:framework} shows the equal-rank numerator; division by the context-constant $N$ only normalizes the score and does not change the ranking. The final influence score is
\begin{equation}
s_{\method}(e)=
\operatorname{mean}\!\left(
\frac{\operatorname{rank}_h(e)}{N},
\frac{\operatorname{rank}_g(e)}{N}
\right).
\label{eq:consensus}
\end{equation}
\method{} sorts $\mathcal C_m$ by Eq.~\eqref{eq:consensus} in decreasing order and resolves exact ties by ascending node index. A high score therefore identifies a candidate that is consistently influential under both readings of the same response. Ranking within each modality prevents different representation spaces from competing directly. The posterior coefficient, propagation steps, and rank weights are fixed across every dataset, predictor, seed, modality, and evaluation depth.

\subsection{Complexity}

\method{} adds no trainable parameters and does not retrain the predictor. For $S$ propagation steps, constructing $Q$ costs $O(S|\cE|C)$ once per dataset and split and is shared by all predictors. The additional posterior storage is $O(|\cV|C)$. Candidate scoring uses one reference-replacement predictor forward; Eqs.~\eqref{eq:hard-score}--\eqref{eq:soft-score} add $O(TC)$ probability operations, where $T=\sum_m\sum_{e=(i,m)\in\mathcal C_m}|\cT_i|$ is the total number of scored candidate--target pairs. Consensus adds $O(N\log N)$ sorting per context.

\begin{figure*}[t]
\centering
\includegraphics[width=0.94\textwidth]{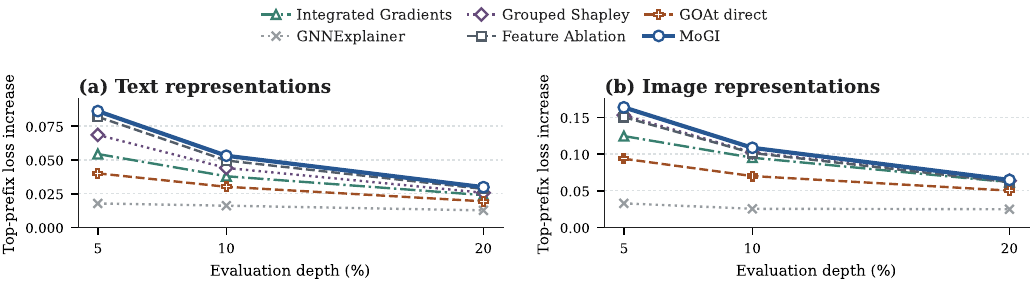}
\caption{\textbf{Modality-resolved comparison.} Mean top-prefix loss increase for all methods under the common protocol. \method{} ranks first for both text and image at every depth.}
\label{fig:modality-main}
\end{figure*}

\section{Experiments}
\label{sec:experiments}

We evaluate whether \method{} identifies influential representations consistently across standard graph neural networks (GNNs) and multimodal graph predictors, how it compares with external attribution methods under the same ranking protocol, and what computation and supervision this task-directed measurement adds.

\subsection{Experimental Settings}

\textbf{Datasets and evaluation.}
The primary comparison uses SemArt, Grocery, Toys, Movies, and RedditS from the MAG benchmarks~\cite{MAGB,OpenMAG}. We evaluate GCN, GraphSAGE, GAT, MMGCN, MGAT, LGMRec, DMGC, and DGF~\cite{GCN,GraphSAGE,GAT,MMGCN,MGAT,LGMRec,DMGC,DGF}, with three seeds and separate text and image rankings. Each dataset--seed--predictor job contains the same frozen 220 candidates, producing 440 candidate--modality scores. The 5\%, 10\%, and 20\% depths inspect the first 11, 22, and 44 representations per modality. All methods share splits, candidates, references, target groups, frozen predictors, and one-at-a-time evaluation. \method{} is training-label-assisted but reads no held-out labels; held-out loss is computed only after every score and ranking has been finalized.

\textbf{Compared methods.}
We compare 32-step Integrated Gradients~\cite{IntegratedGradients}, full-attribute-mask GNNExplainer~\cite{GNNExplainer}, exact two-group Shapley and Feature Ablation~\cite{Captum}, and the reference-direct GOAt term~\cite{GOAt}. The latter is the exact stated local first-order contribution but not a universal full GOAt decomposition for nonlinear predictors; the Supplementary Material separately reports full GOAt on a transparent additive predictor. All learned-predictor methods share candidates, references, graph targets, frozen outputs, and evaluation. A response-only ablation is reported separately in the Supplementary Material because it is an internal component rather than an external method.

\textbf{Reporting and inference.}
Aggregate means average 240 dataset--seed--predictor--modality contexts at each depth. Paired inference first averages predictors and modalities within 15 dataset--seed clusters, then uses 10,000-replicate bootstrap intervals and Holm-corrected Wilcoxon tests; the Supplementary Material reports the complete frozen inferential family.

\begin{table}[t]
\centering
\giveCompactTableSetup
\begin{tabular}{@{}lccc@{}}
\toprule
\givetabhead{Method} & \givetabhead{5\%} & \givetabhead{10\%} & \givetabhead{20\%} \\
\midrule
\multicolumn{4}{c}{\textbf{External attribution methods}} \\
\addlinespace[0.5pt]
Integrated Gradients~\citeyearpar{IntegratedGradients} & \givevalue{0.089522} & \givevalue{0.066516} & \givevalue{0.042869} \\
GNNExplainer~\citeyearpar{GNNExplainer} & \givevalue{0.025325} & \givevalue{0.020792} & \givevalue{0.018749} \\
Grouped Shapley~\citeyearpar{Captum} & \givevalue{0.110854} & \givevalue{0.073264} & \givevalue{0.044687} \\
Feature Ablation~\citeyearpar{Captum} & \givevalue{0.116195} & \givevalue{0.075535} & \givevalue{0.045167} \\
GOAt direct term~\citeyearpar{GOAt} & \givevalue{0.066869} & \givevalue{0.050106} & \givevalue{0.034817} \\
\addlinespace[1.5pt]
\cmidrule(lr){1-4}
\multicolumn{4}{c}{\textbf{Proposed method}} \\
\addlinespace[0.5pt]
\textbf{\method{} (ours)} & \givebestvalue{0.124982} & \givebestvalue{0.080864} & \givebestvalue{0.047339} \\
\bottomrule
\end{tabular}
\caption{\textbf{Learned-predictor comparison.} Aggregate top-prefix loss increase on five datasets, eight predictors, and three seeds. Higher is better; underlines mark best means. GOAt is the qualified reference-direct term.}
\label{tab:main}
\end{table}

\subsection{Influence Across Graph Predictors}

\begin{figure*}[t]
\centering
\includegraphics[width=0.98\textwidth]{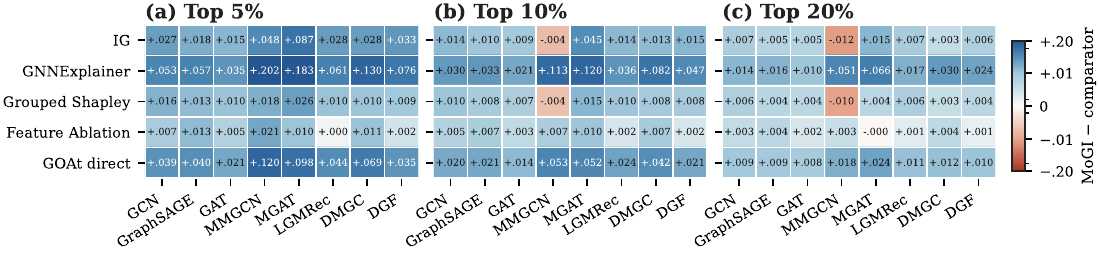}
\caption{\textbf{MoGI margins across graph predictors.} Cells show \method{} minus each comparator; blue, positive values favor \method{}. The symmetric-log color scale expands small margins; GOAt is the qualified reference-direct term.}
\label{fig:predictor-margins}
\end{figure*}

\begin{table*}[t]
\centering
\giveWideTableSetup
\begin{tabular*}{\textwidth}{@{\extracolsep{\fill}}lccccccc@{}}
\toprule
\givetabhead{Graph predictor} & \multicolumn{6}{c}{\givetabhead{\method{} graph-space evaluation}} & \givetabhead{Scoring} \\
\cmidrule(lr){2-7}\cmidrule(lr){8-8}
 & \givetabhead{5\% $\Delta$loss} & \givetabhead{10\% $\Delta$loss} & \givetabhead{20\% $\Delta$loss} & \givetabhead{Lift/random} & \givetabhead{Win rate} & \givetabhead{Min. $\Delta$/random} & \givetabhead{Time (s)} \\
\midrule
\multicolumn{8}{l}{\textbf{Standard GNNs}} \\
GCN & 0.069 & 0.042 & 0.024 & 10.4$\times$ & 82.2\% & \textbf{+0.020} & 7.43 \\
GraphSAGE & 0.076 & 0.043 & 0.023 & 11.7$\times$ & 83.3\% & \textbf{+0.019} & 4.37 \\
GAT & 0.051 & 0.030 & 0.017 & 11.3$\times$ & 82.2\% & \textbf{+0.014} & 16.84 \\
\addlinespace[1.5pt]
\multicolumn{8}{l}{\textbf{Multimodal graph predictors}} \\
MMGCN & 0.251 & 0.150 & 0.082 & 8.2$\times$ & 80.0\% & \textbf{+0.063} & 2.54 \\
MGAT & 0.234 & 0.166 & 0.106 & 7.0$\times$ & 90.0\% & \textbf{+0.081} & 4.73 \\
LGMRec & 0.069 & 0.042 & 0.025 & 11.0$\times$ & 95.6\% & \textbf{+0.020} & 4.28 \\
DMGC & 0.159 & 0.111 & 0.064 & 8.9$\times$ & 96.7\% & \textbf{+0.052} & 1.00 \\
DGF & 0.092 & 0.063 & 0.037 & 8.2$\times$ & 100.0\% & \textbf{+0.030} & 1.41 \\
\bottomrule
\end{tabular*}
\caption{\textbf{Graph-predictor evaluation.} \method{} across graph predictors on five datasets and three seeds. $\Delta$loss is top-prefix loss increase; Lift is the three-depth ratio to matched random, Win rate is the percentage of favorable context--depth rows, and Min. $\Delta$/random is the smallest depth-wise mean gain over random. Time averages scoring 440 candidates.}
\label{tab:graph-main}
\end{table*}

\begin{figure}[t]
\centering
\includegraphics[width=\columnwidth]{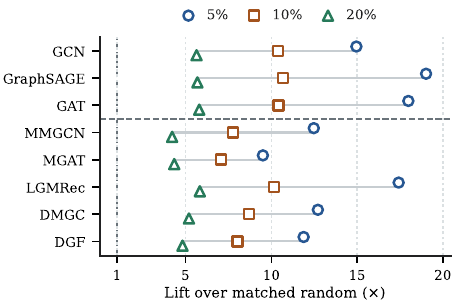}
\caption{\textbf{Lift over matched random.} Ratios are resolved by graph predictor and evaluation depth; the dashed line marks no lift. Every result exceeds $4.2\times$.}
\label{fig:random-lift}
\end{figure}

\textbf{The influence ranking transfers across graph architectures.}
Across all eight predictors, the top-5\% loss increase ranges from $0.051$ to $0.251$, and every three-depth lift over matched random exceeds $7.0\times$ (Table~\ref{tab:graph-main}). Figure~\ref{fig:random-lift} resolves this lift by depth: all 24 ratios exceed $4.2\times$. Figure~\ref{fig:predictor-margins} separately compares every external method. Thus, the graph-space signal is not tied to one predictor family; complete dataset slices appear in the Supplementary Material.

\subsection{Comparison with Attribution Baselines}

\textbf{\method{} establishes the strongest aggregate ranking.}
In the learned-predictor benchmark, Table~\ref{tab:main} and Fig.~\ref{fig:modality-main} show that \method{} ranks first over all five external methods at every aggregate depth and separately for text and image. Feature Ablation is the closest aggregate comparator: the margins are $+0.008787$, $+0.005328$, and $+0.002172$. All have positive clustered 95\% intervals, with Holm-adjusted Wilcoxon $p$-values of $.0092$, $.0092$, and $.0452$. Complete comparisons and internal ablations appear in the Supplementary Material.

\subsection{Efficiency and Ranking Robustness}

\textbf{Task direction adds supervision, not model size.}
\method{} has no learned scorer, trainable parameters, or predictor retraining. Label propagation is computed once per dataset--seed pair and reused across predictors; its largest posterior is 1.42 MiB. Both readings share the replacement forward that produces $P^{-e}$. The 120-job run took 754.9 seconds and peaked at 14,859 MiB allocated GPU memory. Because task direction and rank consensus add no predictor forward, \method{} remains in the one-at-a-time ablation cost class rather than 32-step path integration.

\section{Discussion}

\textbf{What the diagnostic reveals.}
\method{} complements classification accuracy with a representation-level view of model response. Separate modality rankings identify influential text and image inputs, graph-distance groups show whether responses remain local or spread, and predictor comparisons show how architecture changes these patterns. Response magnitude and stability vary across architectures, while every predictor retains positive separation from matched random. The node--modality unit avoids merging modalities into a whole-node score or fragmenting them into non-reviewable coordinates.

Together, Figs.~\ref{fig:modality-main}--\ref{fig:random-lift} show that the advantage persists across external comparators, architecture-specific response scales, and both modalities.

\textbf{Use in model analysis.}
Because the predictor remains frozen, a developer can compare architectures, inspect highly ranked inputs, trace response spread, and prioritize review without modifying the deployed model. Using the same candidates across predictors makes these comparisons directly interpretable: an analyst can examine whether an encoder changes text reliance, whether fusion redistributes it toward images, or whether propagation moves the response farther from the source. The separate modality rankings and graph-distance targets therefore connect architecture choices to concrete representations and response locations, complementing aggregate classification quality.

\section{Conclusion}

We study influence measurement for modality-specific node representations in frozen multimodal graph predictors. \method{} treats each encoded text or image vector as the decision unit, replaces it by a same-modality training reference, and measures task-directed prediction changes at the source and nearby nodes. A training-label posterior supplies direction, while two readings of the same response and their rank consensus produce modality-wise rankings without retraining the predictor. Across five graphs and eight predictors, \method{} outperforms all five external methods at every aggregate depth and remains stronger than matched random across architectures. The rankings reveal how encoder, fusion, and propagation choices shape graph-space reliance. Together, these results establish \method{} as a practical diagnostic for analyzing graph-space model response.

\clearpage
{\small
\bibliography{references}
}

\end{document}


\begin{center}
{\LARGE\bfseries Supplementary Material}
\end{center}
\vspace{0.6em}

This supplement reports the complete empirical results, formal properties, full scoring algorithm, and implementation details for \method{} (Modality-specific Graph Influence). The presentation follows the method's single technical line: replace one text or image representation, read the resulting graph response in the direction of training-label evidence, and combine the two directed readings by rank. Up/down arrows indicate whether higher or lower values are preferred; unmarked quantities are descriptive diagnostics.

\appendix
\begin{center}
{\large\bfseries Contents}
\end{center}
{\footnotesize
\noindent\begin{tabular*}{\columnwidth}{@{}>{\raggedright\arraybackslash}p{0.84\columnwidth}@{\extracolsep{\fill}}r@{}}
A. More Experiments & \pageref{sec:complete-results} \\
B. Theoretical Analysis and Proofs & \pageref{sec:theory} \\
C. Algorithm and Full Scoring Procedure & \pageref{sec:complete-method} \\
D. Implementation and Evaluation Details & \pageref{sec:implementation} \\
E. Environment & \pageref{sec:environment} \\
F. Dataset Details & \pageref{sec:datasets} \\
G. Baseline Details & \pageref{sec:baselines} \\
H. Hyperparameter Details & \pageref{sec:settings} \\
I. Limitations and Deployment Considerations & \pageref{sec:limitations} \\
\end{tabular*}
}

\section{More Experiments}
\label{sec:complete-results}

\subsection{Complete Learned-Predictor Comparison}

Table~\ref{tab:main} reports every learned-predictor comparator under the same candidates, graph targets, and ranking utility.
\begin{center}
\giveCompactTableSetup
\begin{tabular}{@{}lccc@{}}
\toprule
\givetabhead{Method} & \givetabhead{5\%} & \givetabhead{10\%} & \givetabhead{20\%} \\
\midrule
\multicolumn{4}{c}{\textbf{External attribution methods}} \\
\addlinespace[0.5pt]
Integrated Gradients~\citeyearpar{IntegratedGradients} & 0.089522 & 0.066516 & 0.042869 \\
GNNExplainer~\citeyearpar{GNNExplainer} & 0.025325 & 0.020792 & 0.018749 \\
Grouped Shapley~\citeyearpar{Captum} & 0.110854 & 0.073264 & 0.044687 \\
Feature Ablation~\citeyearpar{Captum} & 0.116195 & 0.075535 & 0.045167 \\
GOAt direct term~\citeyearpar{GOAt} & 0.066869 & 0.050106 & 0.034817 \\
\addlinespace[1.5pt]
\cmidrule(lr){1-4}
\multicolumn{4}{c}{\textbf{Proposed method}} \\
\addlinespace[0.5pt]
\textbf{\method{} (ours)} & \givebestvalue{0.124982} & \givebestvalue{0.080864} & \givebestvalue{0.047339} \\
\bottomrule
\end{tabular}
\captionof{table}{\textbf{Learned-predictor comparison.} Aggregate top-prefix loss increase on five datasets, eight predictors, and three seeds. Higher is better; underlines mark best means. GOAt is the qualified reference-direct term.}
\label{tab:main}
\end{center}

Integrated Gradients, GNNExplainer, grouped Shapley, Feature Ablation, and the qualified GOAt direct term all appear in this comparison. Across five datasets, eight predictors, three seeds, and two modalities, \method{} obtains $0.124982/0.080864/0.047339$ mean top-prefix loss increase at 5\%/10\%/20\%. It exceeds \basemethod{} by $+0.012736/+0.007622/+0.002485$, grouped Shapley by $+0.014129/+0.007600/+0.002652$, and Integrated Gradients by $+0.035460/+0.014348/+0.004470$.

Table~\ref{tab:posterior-statistics} gives paired confidence intervals and multiplicity-adjusted tests for the pre-specified inferential family. This family contains three external methods and the response-only ablation; it was specified before the additional GNNExplainer and GOAt evaluations, whose exact means are reported in Table~\ref{tab:main} without changing the family after seeing their results.
\begin{table*}[!ht]
\centering
\giveTableSetup
\begin{tabular}{llcccc}
\toprule
\givetabhead{Comparator} & \givetabhead{Depth} & \givetabhead{Mean margin} & \givetabhead{Clustered 95\% CI} & \givetabhead{Holm $p$} & \givetabhead{Cluster wins} \\
\midrule
Feature Ablation & 5\% & +0.008787 & $[+0.004771,+0.013544]$ & 0.0092 & 12/15 \\
 & 10\% & +0.005328 & $[+0.002674,+0.008239]$ & 0.0092 & 13/15 \\
 & 20\% & +0.002172 & $[+0.000573,+0.003922]$ & 0.0452 & 11/15 \\
\addlinespace[1.5pt]
Response-only ablation & 5\% & +0.012736 & $[+0.006956,+0.019990]$ & 0.0018 & 14/15 \\
 & 10\% & +0.007622 & $[+0.004240,+0.010949]$ & 0.0081 & 13/15 \\
 & 20\% & +0.002485 & $[+0.000382,+0.004488]$ & 0.0479 & 12/15 \\
\addlinespace[1.5pt]
Grouped Shapley & 5\% & +0.014129 & $[+0.009776,+0.018757]$ & 0.0013 & 14/15 \\
 & 10\% & +0.007600 & $[+0.004689,+0.010327]$ & 0.0049 & 14/15 \\
 & 20\% & +0.002652 & $[+0.000878,+0.004390]$ & 0.0452 & 12/15 \\
\addlinespace[1.5pt]
Integrated Gradients & 5\% & +0.035460 & $[+0.017969,+0.054738]$ & 0.0038 & 13/15 \\
 & 10\% & +0.014348 & $[+0.007316,+0.023316]$ & 0.0007 & 15/15 \\
 & 20\% & +0.004470 & $[+0.001991,+0.007013]$ & 0.0215 & 13/15 \\
\bottomrule
\end{tabular}
\caption{\textbf{Paired aggregate inference.} Complete aggregate comparison for \method{}. Intervals use 10,000 dataset--seed cluster bootstrap replicates after averaging predictors and modalities. Two-sided Wilcoxon tests are Holm-adjusted across four comparators and three depths.}
\label{tab:posterior-statistics}
\end{table*}

Every dataset--seed clustered bootstrap interval is positive. Feature Ablation is the aggregate runner-up and supplies the tightest comparison; all three pre-specified Holm-adjusted tests remain below $.05$ under this clustering. At 5\%, the harmful-selection rate---the fraction of selected candidates with negative held-out loss increase---is $18.30\%$ for \method{} and $23.64\%$ for the runner-up. A stricter dataset-only sensitivity analysis is reported below and does not cross the conventional $.05$ threshold with only five datasets.

\begin{figure*}[t]
\centering
\includegraphics[width=0.90\textwidth]{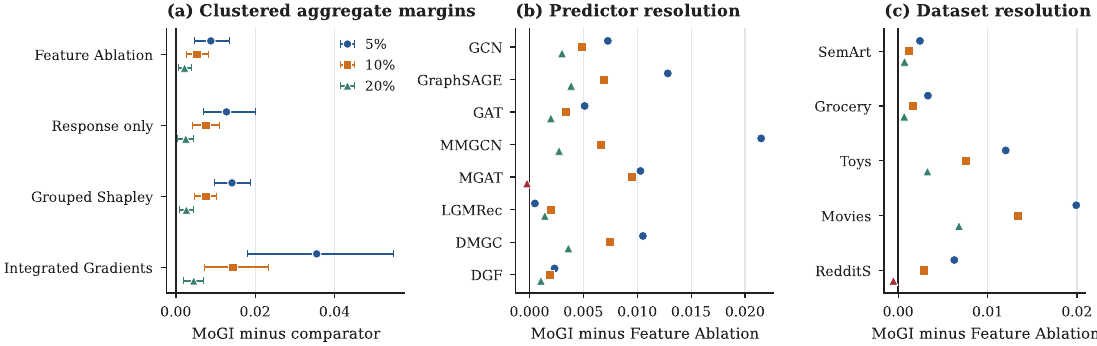}
\caption{\textbf{Aggregate robustness and heterogeneity.} (a) Dataset--seed clustered 95\% bootstrap intervals for \method{} minus each primary comparator. (b--c) Margins over the closest aggregate comparator, Feature Ablation, resolved by predictor and dataset. Red markers denote the two small negative 20\% cells; all aggregate intervals remain positive.}
\label{fig:supp-robustness}
\end{figure*}

Figure~\ref{fig:supp-robustness} separates aggregate uncertainty from predictor- and dataset-level heterogeneity.

Panel (a) supports the aggregate comparison. Panels (b) and (c) expose heterogeneity rather than hiding it inside the mean. The only negative means are MGAT at 20\% and RedditS at 20\%; neither changes the sign of the aggregate or any leave-one-predictor result.

\subsection{Predictor- and Dataset-Resolved Results}

Because Feature Ablation is the aggregate runner-up, Tables~\ref{tab:predictor-main} and~\ref{tab:dataset-margins} resolve this closest comparison. The complete aggregate comparison across every method remains in Table~\ref{tab:main}.
\begin{table}[!ht]
\centering
\giveCompactTableSetup
\begin{tabular}{lrrr}
\toprule
\givetabhead{Predictor} & \givetabhead{5\%} & \givetabhead{10\%} & \givetabhead{20\%} \\
\midrule
\multicolumn{4}{l}{\textbf{Standard GNNs}} \\
GCN & +0.0073 & +0.0048 & +0.0030 \\
GraphSAGE & +0.0128 & +0.0069 & +0.0038 \\
GAT & +0.0051 & +0.0034 & +0.0020 \\
\addlinespace[1.5pt]
\multicolumn{4}{l}{\textbf{Multimodal graph predictors}} \\
MMGCN & +0.0215 & +0.0066 & +0.0027 \\
MGAT & +0.0103 & +0.0095 & -0.0002 \\
LGMRec & +0.0005 & +0.0020 & +0.0014 \\
DMGC & +0.0105 & +0.0074 & +0.0036 \\
DGF & +0.0023 & +0.0019 & +0.0010 \\
\bottomrule
\end{tabular}
\caption{\textbf{Predictor-resolved margin.} \method{} minus Feature Ablation in mean top-prefix loss increase. Positive values favor \method{}; the single negative predictor--depth mean remains visible.}
\label{tab:predictor-main}
\end{table}

\method{} exceeds Feature Ablation for all eight predictor means at 5\% and 10\%, and seven of eight at 20\%. The only negative predictor--depth mean is MGAT at 20\% ($-0.000238$). Removing any one predictor leaves a positive aggregate margin at all depths; the minimum leave-one-predictor margins are $+0.006971$, $+0.004728$, and $+0.001932$.

\begin{table}[!ht]
\centering
\giveCompactTableSetup
\begin{tabular}{lrrr}
\toprule
\givetabhead{Dataset} & \givetabhead{5\%} & \givetabhead{10\%} & \givetabhead{20\%} \\
\midrule
SemArt & +0.0024 & +0.0012 & +0.0007 \\
Grocery & +0.0033 & +0.0016 & +0.0007 \\
Toys & +0.0120 & +0.0076 & +0.0033 \\
Movies & +0.0199 & +0.0133 & +0.0068 \\
RedditS & +0.0063 & +0.0029 & -0.0006 \\
\bottomrule
\end{tabular}
\caption{\textbf{Dataset-resolved margin.} \method{} minus Feature Ablation by dataset. Positive values favor \method{}; RedditS at 20\% is the only negative dataset--depth mean.}
\label{tab:dataset-margins}
\end{table}

\method{} exceeds Feature Ablation on every dataset mean at 5\% and 10\%, and on four of five at 20\%. RedditS at 20\% is the only negative dataset--depth mean ($-0.000553$). These two small exceptions prevent a literal clean-sweep claim.

\subsection{Validation-Based Method Selection}
\label{sec:validation-selection}

Table~\ref{tab:posterior-components} compares the two directed readings, their rank consensus, and the strongest response baselines on held-out data.
\begin{table*}[!ht]
\centering
\giveTableSetup
\begin{tabular}{@{}lcccc@{}}
\toprule
\givetabhead{Ranking} & \givetabhead{5\%} & \givetabhead{10\%} & \givetabhead{20\%} & \givetabhead{Harmful at 5\% ($\downarrow$)} \\
\midrule
Feature Ablation & \givevalue{0.116195} & \givevalue{0.075535} & \givevalue{0.045167} & \givevalue{23.64\%} \\
Response-only ablation & \givevalue{0.112246} & \givevalue{0.073241} & \givevalue{0.044853} & \givevalue{33.75\%} \\
Posterior-targeted drop & \givevalue{0.121505} & \givevalue{0.079507} & \givevalue{0.047376} & \givevalue{19.85\%} \\
Posterior expected log-loss & \givevalue{0.126384} & \givevalue{0.082310} & \givevalue{0.048372} & \givevalue{23.60\%} \\
\textbf{\method{} (ours)} & \givebestvalue{0.124982} & \givebestvalue{0.080864} & \givebestvalue{0.047339} & \givebestvalue{18.30\%} \\
\bottomrule
\end{tabular}
\caption{\textbf{Posterior component diagnostic.} Held-out ranking utility and harmful-selection rate for the two directed readings, their equal-rank consensus, and response baselines.}
\label{tab:posterior-components}
\end{table*}

The final method was selected before held-out evaluation. On independent validation seeds 45 and 46, equal-rank consensus exceeded Feature Ablation by $+0.005784/+0.002730/+0.002017$, won on seven of eight predictor means and all five dataset means at each depth, and reduced 5\% harmful selection from $23.46\%$ to $18.61\%$. The same rule was then used unchanged for held-out evaluation.

The soft posterior expected-log-loss constituent is descriptively higher than consensus on the final test aggregate. It did not satisfy the pre-test cross-predictor validation criterion and was therefore not selected post hoc. It is reported only as a diagnostic; equal-rank consensus remains the scoring rule used by \method{}.

\begin{figure*}[t]
\centering
\includegraphics[width=0.98\textwidth]{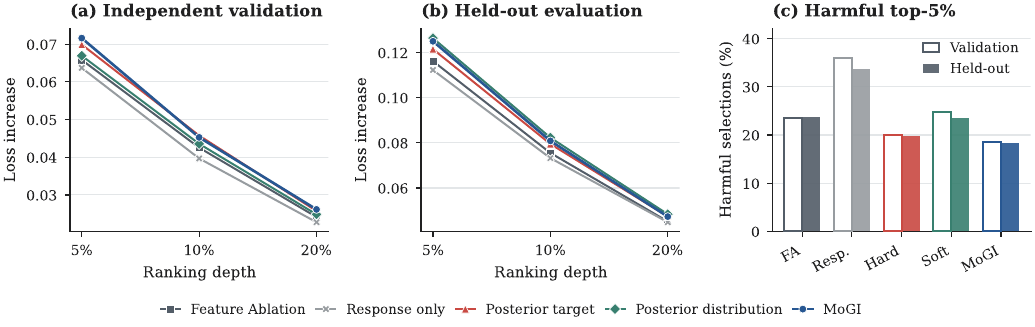}
\caption{\textbf{Validation-based method selection.} (a) Component results on independent validation seeds, where equal-rank consensus met the pre-test criterion. (b) Held-out results obtained without changing the selected rule. (c) Harmful top-5\% selections on validation and held-out data. The soft constituent is a diagnostic, not a post-hoc replacement for \method{}.}
\label{fig:supp-selection}
\end{figure*}

Figure~\ref{fig:supp-selection} displays validation and held-out results side by side so that this chronology is explicit.

Parameter selection used validation data only. On development seed 44, the posterior mixture was checked at $\{0,.10,.25,.50,.75,1\}$; $.10$ best met the prespecified cross-predictor, dataset, harmful-selection, and leave-one-predictor criteria. Hard-rank weights from $.30$ through $.95$ also met the diagnostic gate, so equal weighting was fixed as the central symmetric choice rather than at a narrow optimum. The complete formula was then evaluated without alteration on independent validation seeds 45 and 46.

\subsection{Transparent-Predictor Attribution Comparison}

Table~\ref{tab:transparent-goat} places all eight available rankings together on a transparent additive predictor, where the full GOAt decomposition is exact.
\begin{table}[!ht]
\centering
\giveCompactTableSetup
\begin{tabular}{@{}lccc@{}}
\toprule
\givetabhead{Method} & \givetabhead{5\%} & \givetabhead{10\%} & \givetabhead{20\%} \\
\midrule
Random & \givevalue{0.022050} & \givevalue{0.022103} & \givevalue{0.022214} \\
Grouped Shapley~\citeyearpar{Captum} & \givevalue{0.102579} & \givevalue{0.084177} & \givevalue{0.062699} \\
Integrated Gradients~\citeyearpar{IntegratedGradients} & \givevalue{0.102314} & \givevalue{0.086833} & \givevalue{0.065944} \\
GNNExplainer~\citeyearpar{GNNExplainer} & \givevalue{0.017299} & \givevalue{0.020644} & \givevalue{0.023381} \\
Feature Ablation~\citeyearpar{Captum} & \givevalue{0.111895} & \givevalue{0.086239} & \givevalue{0.064379} \\
GOAt~\citeyearpar{GOAt} & \givevalue{0.108287} & \givevalue{0.089220} & \givevalue{0.066967} \\
Response-only ablation & \givebestvalue{0.121331} & \givebestvalue{0.090561} & \givebestvalue{0.068103} \\
\textbf{\method{} (ours)} & \givevalue{0.103910} & \givevalue{0.085856} & \givevalue{0.064638} \\
\bottomrule
\end{tabular}
\caption{\textbf{Transparent-predictor comparison.} Complete attribution comparison on the held-fixed transparent additive predictor over five datasets, three seeds, and two modalities. GOAt is exact in this setting because the predictor is linear before softmax; this exact decomposition is not extrapolated to the eight learned predictors. Higher top-prefix loss increase is better.}
\label{tab:transparent-goat}
\end{table}

The predictor, candidates, targets, and held-out loss-increase evaluation are held fixed across methods. The response-only ablation ranks first at all three depths, while GOAt ranks third, second, and second and \method{} ranks fourth, fifth, and fourth. This result is reported separately because the transparent predictor differs from the eight learned predictors in the primary benchmark; it is not evidence for a universal clean sweep.

\subsection{Numerical, Statistical, and Integrity Checks}
\label{sec:integrity}

The final run covers $5$ datasets $\times$ $8$ predictors $\times$ $3$ seeds $=120$ jobs. Each job scores 220 candidates in text and image, giving 52,800 candidate--modality rows for each newly computed method. All jobs completed before held-out loss increase was read. The checks verify complete candidate coverage, finite scores, deterministic ranking reconstruction, exact recomputation of original and reference-replaced predictions, total-variation and Feature Ablation replay, grouped-Shapley efficiency, training-label-only propagation, and exact tie ordering.

In all 15 dataset--seed posterior records, the validation- and test-label indexing flags are false, nontraining rows are masked before propagation, and posterior rows sum to one within $3.58\times10^{-7}$. Test split indices identify evaluation targets, but their labels are not used for scoring. Scores and rankings are saved before held-out utility is read, and an independent reconstruction reproduces the final rankings from raw scores.

Every learned-predictor method uses the same trained predictor, candidate nodes, representations, training-mean references, and evaluation targets. We also checked Integrated Gradients at 16, 32, and 64 steps on Grocery/GCN/seed 44. Every audited within-modality ranking has Spearman correlation $1.000$. From 32 to 64 steps, the mean and maximum absolute score changes are $3.45\times10^{-5}$ and $4.07\times10^{-4}$, supporting the pre-specified 32-step setting.

The pre-specified uncertainty unit is a dataset--seed group after averaging predictors and modalities, yielding 15 clusters rather than 240 context rows treated as independent. Confidence intervals use 10,000 cluster-bootstrap replicates. Two-sided Wilcoxon tests use the same paired cluster means, and Holm correction covers four comparators at three depths. Predictor, dataset, modality, seed, and leave-one-predictor summaries are retained separately.

Seeds on the same graph are repeated training realizations rather than independent datasets. We therefore add a conservative sensitivity analysis that first averages over seeds, predictors, and modalities and uses only the five datasets as paired units. Against Feature Ablation, all five dataset margins are positive at 5\% and 10\%, and four of five are positive at 20\%. Exact two-sided Wilcoxon $p$-values are $.0625$, $.0625$, and $.125$, respectively. Thus the direction of the aggregate result is stable, but five-dataset inference alone does not support a conventional $p<.05$ claim.

\subsection{Complexity, Runtime, and Memory}
\label{sec:complexity}

\method{} adds no trainable parameters. Label propagation costs $O(S|\mathcal E|C)$ once per dataset--seed pair and is reused across all eight predictors. The posterior matrix requires $O(|\mathcal V|C)$ float32 storage; one posterior occupies 0.816--1.421 MiB across the five datasets.

Hard and soft readings reuse the same original and replacement probabilities as \basemethod{} and Feature Ablation. They add no predictor forward per candidate--modality pair. The final run required 587.32 seconds in the shared score loop and 754.95 seconds end to end, including model loading, 15 posterior constructions, score computation, held-out loss parsing, and aggregation. Peak allocated GPU memory was 14,859 MiB on GAT; peak resident CPU memory was 2,978 MiB.

\begin{figure*}[t]
\centering
\includegraphics[width=0.98\textwidth]{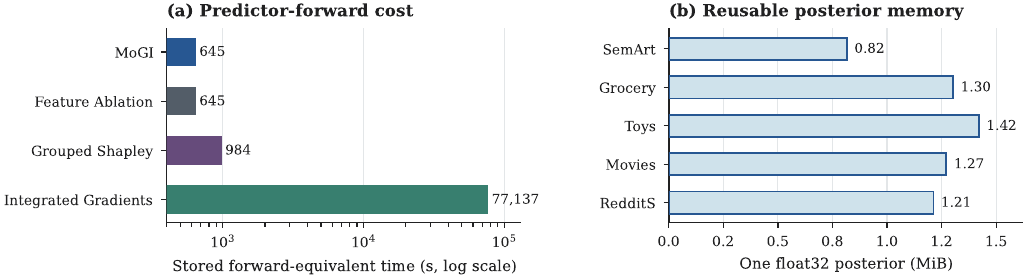}
\caption{\textbf{Cost boundary.} (a) Saved forward-equivalent predictor time over 120 jobs. Shared coalition forwards are credited to each method that requires them; the values are not standalone end-to-end wall times. \method{} and Feature Ablation require the same one-at-a-time replacement forwards. (b) Float32 storage for one reusable training-label posterior.}
\label{fig:supp-efficiency}
\end{figure*}

The shared loop materializes several response scores, so its wall time is not a standalone per-method comparison. Summed predictor-forward durations are 644.53 seconds for \method{}, \basemethod{}, and Feature Ablation, 983.81 seconds for grouped Shapley, and 77,136.64 seconds for 32-step Integrated Gradients. Figure~\ref{fig:supp-efficiency} separates these forward-equivalent times from posterior storage.

\section{Theoretical Analysis and Proofs}
\label{sec:theory}

Let $e=(i,m)$ denote the modality-$m$ representation of node $i$, let $P_j$ and $P_j^{-e}$ be the original and reference-replaced predictions at target $j$, and let $Q_j$ be the distribution propagated from training labels. The target set $\cT_i$ contains evaluation nodes at distances zero, one, and two from $i$. The following properties state what the reference, posterior direction, graph-response readings, and rank consensus guarantee. They do not make a causal claim or assert completeness outside the stated graph scope.

\subsection{Reference and Supervision Boundary}

\begin{proposition}[Training-mean reference]
For a nonempty training set and modality $m$, the reference $r^m=|\mathcal V_{\mathrm{tr}}|^{-1}\sum_{k\in\mathcal V_{\mathrm{tr}}}h_k^m$ uniquely minimizes $\sum_{k\in\mathcal V_{\mathrm{tr}}}\|h_k^m-z\|_2^2$ over vectors $z$ in the modality representation space.
\end{proposition}
\begin{proof}
Write $n=|\mathcal V_{\mathrm{tr}}|$ and $r=r^m$. Since $\sum_k(h_k^m-r)=0$,
\[
\sum_k\|h_k^m-z\|_2^2=\sum_k\|h_k^m-r\|_2^2+n\|z-r\|_2^2.
\]
The second term is nonnegative and vanishes only at $z=r$.
\end{proof}

The reference is the least-squares representative of training embeddings in that modality. It is not a semantic absence operator or a claim of causal removal.

\begin{proposition}[Held-out-label invariance]
Fix the graph, training-node set, propagation parameters, and number of iterations. If two complete label assignments agree on every training node, the posterior $Q$ produced by clamped propagation is identical under both assignments, regardless of their validation or test labels.
\end{proposition}
\begin{proof}
The seed matrix $Y$ keeps the common training labels and sets every nontraining row to zero, so it is identical for both assignments. Hence $Q^{(0)}=Y$ is identical. If $Q^{(t)}$ is identical, multiplication by the same $\widetilde A$, addition of the same restart term, and clamping to the same training rows produce identical $Q^{(t+1)}$. Induction and deterministic row normalization preserve equality.
\end{proof}

Held-out node identities may define evaluation targets, but their label values do not construct $Q$, a score, or a ranking.

\subsection{Task Direction and Graph-Response Readings}

\begin{proposition}[Bounded posterior direction]
If $P_j$ and $Q_j$ are class distributions and $\beta\in[0,1]$, then $R_j=(1-\beta)P_j+\beta Q_j$ is a class distribution and $\TV(R_j,P_j)=\beta\,\TV(Q_j,P_j)\leq\beta$.
\end{proposition}
\begin{proof}
$R_j$ is a convex combination of two simplex points. Since $R_j-P_j=\beta(Q_j-P_j)$, homogeneity of the $\ell_1$ norm gives the equality. Total variation between two distributions is at most one.
\end{proof}

Training-label evidence therefore supplies a controlled direction rather than replacing the predictor distribution.

\begin{proposition}[Meaning of the hard and soft readings]
Define $a(e)=|\cT_i|^{-1}\sum_{j\in\cT_i}\TV(P_j,P_j^{-e})$. Then the hard reading satisfies $|h(e)|\leq a(e)$. If all class probabilities are positive, the soft reading satisfies
\[
g(e)=\frac{1}{|\cT_i|}\sum_{j\in\cT_i}
\left[\KL(R_j\|P_j^{-e})-\KL(R_j\|P_j)\right].
\]
\end{proposition}
\begin{proof}
For any class $c$, $|P_j(c)-P_j^{-e}(c)|$ is at most $\TV(P_j,P_j^{-e})$. The triangle inequality after averaging gives the hard-score bound. For the soft reading, expand both Kullback--Leibler divergences. Their $\sum_cR_j(c)\log R_j(c)$ terms cancel, leaving $\sum_cR_j(c)[\log P_j(c)-\log P_j^{-e}(c)]$.
\end{proof}

The hard reading tracks the posterior-preferred class; the soft reading tracks the full distribution. Both are signed, so negative movement is retained. Log arguments are clipped only at machine-scale epsilon for numerical safety.

\subsection{Rank Consensus}

\begin{proposition}[Properties of equal-rank consensus]
Let $u_e$ and $v_e$ be the normalized average-tie ranks of $h(e)$ and $g(e)$. The score $s(e)=(u_e+v_e)/2$ (i) uniquely minimizes $(z-u_e)^2+(z-v_e)^2$ over $z\in\mathbb R$; (ii) is unchanged by a strictly increasing rescaling of either reading; and (iii) preserves any strict pairwise preference shared by both readings.
\end{proposition}
\begin{proof}
Differentiating the quadratic gives the unique minimizer $z=(u_e+v_e)/2$. A strictly increasing transformation preserves order and exact ties, hence preserves average-tie ranks. If both rank differences are nonnegative and at least one is positive, their sum is positive.
\end{proof}

The equal average is therefore a symmetric, scale-independent compromise between the two orderings.

\subsection{When a Finite Graph Radius Is Exact}

\begin{proposition}[Local message-passing support]
Consider an $L$-layer predictor in which each layer updates a node using only that node and its one-hop neighbors, with fixed topology and no global operation. Replacing a representation at node $i$ cannot change the output at node $j$ when their shortest-path distance exceeds $L$.
\end{proposition}
\begin{proof}
After one layer, only $i$ and its one-hop neighbors can depend on the changed input. If the affected set after layer $\ell$ lies within $\ell$ hops of $i$, one more local update expands it by at most one hop. Induction gives support within $L$ hops after $L$ layers.
\end{proof}

Global propagation, hypergraph operations, or input-derived structures can violate this premise. The benchmark therefore treats the self/one-hop/two-hop union as a fixed empirical scope, not a universal completeness claim.

\section{Algorithm and Full Scoring Procedure}
\label{sec:complete-method}

\subsection{One Reference-Replacement Response}

Let $\mathcal V$ be the graph nodes, $C$ the number of classes, and $\mathcal V_{\mathrm{tr}}$ the training-node set. For modality $m$, the reference is $r^m=|\mathcal V_{\mathrm{tr}}|^{-1}\sum_{k\in\mathcal V_{\mathrm{tr}}}h_k^m$. Candidate $e=(i,m)$ is the modality-$m$ vector at node $i$. Scoring replaces only this vector by $r^m$ and leaves all other inputs and model parameters unchanged. The trained predictor produces original probabilities $P$ and reference-replaced probabilities $P^{-e}$.

The matrix $Y\in\{0,1\}^{|\mathcal V|\times C}$ contains one-hot labels on $\mathcal V_{\mathrm{tr}}$ and zeros elsewhere. Starting from $Q^{(0)}=Y$, propagation applies
\[
\begin{aligned}
\overline Q^{(t+1)}&=\alpha Y+(1-\alpha)\widetilde A Q^{(t)},\\
Q_i^{(t+1)}&=
\begin{cases}
Y_i,&i\in\mathcal V_{\mathrm{tr}},\\
\overline Q_i^{(t+1)},&i\notin\mathcal V_{\mathrm{tr}},
\end{cases}
\end{aligned}
\]
where $\widetilde A$ is row-normalized adjacency. Final rows are normalized; a row with no propagated mass becomes uniform. Validation and test labels are not used. The task-directed distribution is $R_j=(1-\beta)P_j+\beta Q_j$.

Target nodes are partitioned into mutually exclusive self, one-hop, and two-hop shells whose union is $\cT_i$. For $c_j=\arg\max_cR_j(c)$, the two readings are
\[
h(e)=\frac{1}{|\cT_i|}\sum_{j\in\cT_i}\left[P_j(c_j)-P_j^{-e}(c_j)\right],
\]
\[
g(e)=\frac{1}{|\cT_i|}\sum_{j\in\cT_i}\sum_cR_j(c)
\left[\log P_j(c)-\log P_j^{-e}(c)\right].
\]
Within each dataset--seed--predictor--modality context, larger readings receive larger average-tie ranks. For $N$ candidates,
\[
s(e)=\frac{\operatorname{rank}_h(e)+\operatorname{rank}_g(e)}{2N}.
\]
Candidates are ordered by decreasing $s(e)$, with exact ties resolved by ascending node index. Text and image representations are ranked separately.

\subsection{Pseudocode}

Algorithm~\ref{alg:mogi} makes the data boundary and shared response explicit. It adds no learned parameters and never updates the graph predictor.
\begin{algorithm*}[!t]
\DontPrintSemicolon
\SetAlgoLined
\KwIn{Encoded graph $(A,\{H^m\}_{m\in\cM})$; trained predictor $f_\theta$; training nodes and labels $\mathcal V_{\mathrm{tr}},Y$; candidate sets $\{\mathcal C_m\}$; evaluation-target identities $\mathcal V_{\mathrm{eval}}$; propagation and shell parameters.}
\KwOut{Influence score $s(e)$ for each candidate and one ranking for each modality.}
\tcp{Training-label direction}
Compute the training-mean reference $r^m$ for each modality $m$\;
Build $Q$ by fixed-step propagation from $Y$, clamping training rows after every step\;
Compute original predictions $P\leftarrow f_\theta(A,\{H^m\})$\;
\tcp{One-at-a-time graph responses}
\ForEach{$m\in\cM$}{
  \ForEach{$e=(i,m)\in\mathcal C_m$}{
    Build disjoint self/one-hop/two-hop targets $\cT_i\subseteq\mathcal V_{\mathrm{eval}}$\;
    Replace $h_i^m$ by $r^m$ and compute $P^{-e}$ with the same $f_\theta$\;
    Form $R_j=(1-\beta)P_j+\beta Q_j$ for $j\in\cT_i$\;
    Compute $h(e)$ and $g(e)$ from $P$, $P^{-e}$, and $R$\;
  }
  \tcp{Scale-independent consensus within one modality}
  Assign average-tie ranks to $h$ and $g$, then combine them into $s$\;
  Sort $\mathcal C_m$ by decreasing $s$, then by ascending node index\;
}
\caption{Modality-specific Graph Influence (\method{}).}
\label{alg:mogi}
\end{algorithm*}

\section{Implementation and Evaluation Details}
\label{sec:implementation}

\subsection{Candidate and Target Construction}

Each dataset--seed pair uses a stored array of 220 candidate nodes. The array was produced by seeded, label-blind sampling from nodes with at least one evaluation target; eligibility depends only on graph topology and split membership. The same array is reused across all eight predictors and all ranking methods. Each candidate contributes one text representation and one image representation, which are scored and ranked separately.

For a candidate node, self, one-hop, and two-hop targets are restricted to the evaluation split and made mutually exclusive. Nodes in each shell are sorted by node index and the first $B$ are retained; self contains at most the candidate itself. We use $B=120$. Every retained candidate has a nonempty target union. This rule uses neither labels nor predictions and is identical for all methods.

\subsection{Held-Out Ranking Utility}

Held-out labels evaluate a completed ranking but do not construct it. For candidate $e=(i,m)$, its local loss increase is
\[
u(e)=\frac{1}{|\cT_i|}\sum_{j\in\cT_i}
\left[\CE(P_j^{-e},y_j)-\CE(P_j,y_j)\right].
\]
A positive value means that replacing the representation makes held-out prediction worse. For depth $d$, let $k=\min(N,\max(5,\lceil dN\rceil))$ and let $\pi_m$ be the modality-$m$ ranking. The reported utility is
\[
U_d(\pi_m)=\frac{1}{k}\sum_{q=1}^{k}u(\pi_m(q)).
\]
With $N=220$, the 5\%, 10\%, and 20\% depths select 11, 22, and 44 representations per modality. The harmful-selection rate is the selected fraction with $u(e)<0$.

\subsection{Reconstruction Interface}

Section~\ref{sec:integrity} reports the reconstruction results. Saved rows are keyed by dataset, predictor, seed, node, and modality and retain raw readings, ranks, and evaluation utility. Checksums identify the graph, features, split, candidates, checkpoint, and predictions. The archive supports hash audit, table and figure reconstruction, and a self-contained scoring-kernel smoke. Replaying all 120 GPU jobs also requires the public tensors and large frozen checkpoints identified by those checksums but omitted from the size-limited archive.

\section{Environment}
\label{sec:environment}

Experiments use Ubuntu 24.04.3 LTS, two AMD EPYC 9654 96-core processors, 384 GiB RAM, and an NVIDIA RTX PRO 6000 Blackwell Server Edition (96 GiB; driver 580.95.05). The environment contains Python 3.10.18, PyTorch 2.8.0 with CUDA 12.8, DGL 2.4.0, PyTorch Geometric 2.7.0, and Captum 0.8.0. Seeds are 44, 45, and 46; the code package records commands and checksums.

\section{Dataset Details}
\label{sec:datasets}

Table~\ref{tab:datasets} summarizes the five public multimodal graphs from the MAG benchmarks~\cite{MAGB,OpenMAG}.
\begin{table}[t]
\centering
\giveCompactTableSetup
\begin{tabular}{lrrr}
\toprule
\givetabhead{Dataset} & \givetabhead{Nodes} & \givetabhead{Adj. nonzeros} & \givetabhead{Classes} \\
\midrule
RedditS & 15,894 & 283,080 & 20 \\
Movies & 16,672 & 160,802 & 20 \\
Grocery & 17,074 & 142,262 & 20 \\
Toys & 20,695 & 113,402 & 18 \\
SemArt & 21,382 & 1,173,014 & 10 \\
\bottomrule
\end{tabular}
\caption{\textbf{Dataset statistics.} Adjacency nonzeros are counted after removing self loops and symmetrizing the graph.}
\label{tab:datasets}
\end{table}

All five public multimodal node-classification graphs provide text and image representations and fixed 60/20/20 splits. SemArt has 10 artwork types; Grocery, Movies, and Toys have 20, 20, and 18 product categories; RedditS has 20 image--caption post categories.

\section{Baseline Details}
\label{sec:baselines}

\subsection{Graph Predictors}

Table~\ref{tab:predictor-hparams} gives architecture-specific settings, while Table~\ref{tab:predictor-quality} reports the quality of the exact checkpoints used for influence scoring.
\begin{table}[t]
\centering
\small
\setlength{\tabcolsep}{0.8mm}
\renewcommand{\arraystretch}{1.02}
\begin{tabular}{@{}l c >{\raggedright\arraybackslash}p{0.53\columnwidth}@{}}
\toprule
Predictor & $d/L/\mathrm{drop}$ & Architecture-specific setting \\
\midrule
GCN & 256/3/.20 & -- \\
GAT & 256/3/.10 & 3 heads; attention dropout .10 \\
GraphSAGE & 256/3/.02 & Mean aggregation \\
LGMRec & 128/3/.20 & 64 hyperedges; global weight .10; deterministic evaluation \\
DGF & 64/10/-- & $\alpha=1$, $\beta=1$, $\theta=.3$ \\
DMGC & 128/1/.50 & $\tau=1$, $\beta=.1$, $k_l=10$, $k_h=6$ \\
MMGCN & 128/2/-- & Mean aggregation; no ID embedding \\
MGAT & 128/1/-- & No ID embedding \\
\bottomrule
\end{tabular}
\caption{\textbf{Predictor settings.} The compact column gives width $d$, layers $L$, and dropout. Predictors use Adam with learning rate $10^{-3}$, weight decay $5\times10^{-4}$, at most 200 epochs, and patience 20.}
\label{tab:predictor-hparams}
\end{table}

All predictors are trained under the shared node-classification objective and then held fixed. Candidate scoring retains each architecture's nonlinear fusion and propagation. LGMRec uses deterministic zero-noise evaluation, DMGC keeps its full-input feature-derived graphs fixed during scoring, and MMGCN and MGAT omit node-ID embeddings under the common classification protocol.

\begin{table}[t]
\centering
\giveCompactTableSetup
\begin{tabular}{lcc}
\toprule
\givetabhead{Predictor} & \givetabhead{Macro-F1 ($\uparrow$)} & \givetabhead{Accuracy ($\uparrow$)} \\
\midrule
GCN & $0.613\pm0.182$ & $0.678\pm0.195$ \\
GraphSAGE & $0.646\pm0.162$ & $0.703\pm0.209$ \\
GAT & $0.588\pm0.178$ & $0.663\pm0.183$ \\
MMGCN & $0.687\pm0.148$ & $0.737\pm0.177$ \\
MGAT & $\underline{0.711}\pm0.150$ & $\underline{0.755}\pm0.182$ \\
LGMRec & $0.639\pm0.171$ & $0.702\pm0.198$ \\
DMGC & $0.655\pm0.169$ & $0.708\pm0.201$ \\
DGF & $0.698\pm0.164$ & $0.752\pm0.182$ \\
\bottomrule
\end{tabular}
\caption{\textbf{Predictor quality.} Mean $\pm$ population standard deviation over five datasets and three seeds under the shared node-classification protocol. Underlines mark exact best means.}
\label{tab:predictor-quality}
\end{table}

Every architecture has nontrivial held-out macro-F1 and accuracy under the shared task. These values verify the checkpoints used for scoring, not reproduce each architecture's original task-specific result, whose datasets, objectives, and tuning budgets differ. The common protocol changes the graph predictor while holding the influence task, representations, splits, candidates, and evaluation fixed.

\subsection{Attribution Methods}

Feature Ablation uses predicted-class probability drop; Integrated Gradients follows the path from the training mean to the candidate; grouped Shapley enumerates the four text--image coalitions; and GNNExplainer optimizes a full attribute mask. The response-only ablation averages target-wise total variation. Learned-predictor GOAt is the qualified reference-direct clean-point term; its exact full decomposition is confined to the transparent additive predictor. Methods share the endpoint, not their attribution semantics.

\section{Hyperparameter Details}
\label{sec:settings}

Table~\ref{tab:hyperparameters} separates the method and shared-protocol settings from comparator-specific configurations. The shared rows define the common evaluation interface, while the continuation retains each comparator's attribution semantics.
\begin{table}[t]
\centering
\small
\setlength{\tabcolsep}{1.8pt}
\renewcommand{\arraystretch}{1.06}
\begin{tabular}{@{}>{\raggedright\arraybackslash}p{0.34\columnwidth}>{\raggedright\arraybackslash}p{0.59\columnwidth}@{}}
\toprule
\givetabhead{Component} & \givetabhead{Setting} \\
\midrule
Reference & Training-node mean within each modality \\
Training-label propagation & 5 steps; restart $\alpha=.15$; clamped seeds \\
Posterior mixture & $\beta=.10$ \\
Consensus & Hard/soft rank weights $.50/.50$ \\
Targets & At most 120 per nonempty distance shell \\
Candidates & 220 nodes; text and image scored separately \\
Tie handling & Average ranks for ties; ascending node index as the final tie-break \\
Evaluation prefixes & 5\%/10\%/20\%; 11/22/44 entries per modality \\
\bottomrule
\end{tabular}
\caption{\textbf{Hyperparameter settings.} MoGI and shared evaluation-protocol configurations.}
\label{tab:hyperparameters}
\end{table}

\begin{table}[t]
\ContinuedFloat
\centering
\small
\setlength{\tabcolsep}{1.8pt}
\renewcommand{\arraystretch}{1.06}
\begin{tabular}{@{}>{\raggedright\arraybackslash}p{0.34\columnwidth}>{\raggedright\arraybackslash}p{0.59\columnwidth}@{}}
\toprule
\givetabhead{Comparator} & \givetabhead{Setting} \\
\midrule
Integrated Gradients & 32 integration steps \\
Grouped Shapley (learned) & Exact two-modality coalitions \\
Grouped Shapley (transparent) & 32 Captum samples \\
Feature Ablation & Predicted-class probability drop \\
\basemethod{} & Mean target-wise total variation \\
GNNExplainer (learned) & Full attribute mask; 50 epochs; learning rate .05; size/entropy $.01/.01$ \\
GOAt direct term (learned) & Absolute reference-relative clean-point gradient contribution \\
GOAt (transparent) & Exact additive contributions to pre-softmax logits relative to the shared reference \\
\bottomrule
\end{tabular}
\caption{\textbf{Comparator settings.} Learned GNNExplainer uses a full attribute mask; exact full GOAt is confined to the transparent additive predictor.}
\end{table}

Section~\ref{sec:validation-selection} documents the selection chronology and leakage checks. The resulting settings are shared across datasets, predictors, seeds, modalities, candidates, and evaluation depths.

\section{Limitations and Deployment Considerations}
\label{sec:limitations}

\paragraph{Supervision and graph availability.}
\method{} is training-label-assisted, not label-free. It needs the training split and graph, and its posterior can reflect label noise, imbalance, or weak labeled connectivity. Validation and test label values remain unused. The external comparison matches predictors, candidates, references, targets, and endpoints but is not information-matched because most baselines do not use this posterior.

\paragraph{Reference and score meaning.}
The training mean is a least-squares representative, not semantic absence. A high score records predictor change over the stated targets; it does not establish data quality, human interpretability, or causality. Rank consensus is scale-robust but discards magnitude spacing and fixes equal weights.

\paragraph{Representation dependence.}
Responses are conditional on encoded features and the checkpoint, so changing the encoder, fusion, or propagation requires recomputation. Scores are not calibrated across predictors or modalities: quantitative cross-architecture reliance claims need separate calibration or rank-overlap analysis.

\paragraph{Scope and coverage.}
The benchmark covers text and image, 220 candidates, capped two-hop targets, and node classification on five graphs. The locality result requires strictly local message passing; global or input-derived components need not be two-hop complete. Other modalities, tasks, references, and inductive graphs remain outside the evidence.

\paragraph{Comparison boundary.}
It is not a universal claim. MGAT and RedditS are below Feature Ablation at 20\%, although the aggregate margin and dataset--seed interval are positive; dataset-only sensitivity preserves direction but not conventional $.05$ significance. Exact full GOAt is confined to the transparent additive setting, where response-only ablation leads. Different semantics remain under the matched endpoint, so the advantage cannot be assigned solely to consensus or one \method{} component.

\paragraph{Singleton evaluation and cost.}
The benchmark evaluates one-at-a-time replacement and prefixes, not jointly replaced subsets or subset optimality. Forward-equivalent times are hardware-specific. Deployment should treat \method{} as a trained-predictor diagnostic and retain predictor-, dataset-, modality-, and graph-regime checks.

{\footnotesize
\setlength{\bibsep}{1pt}
\bibliography{references}
}